\documentclass[10pt,a4paper,oneside]{article}
\usepackage{amssymb}
\usepackage{amsmath}
\usepackage{graphicx}
\usepackage{theorem}
\usepackage{setspace}
\usepackage{appendix}
\usepackage{indentfirst}
\usepackage[noend]{algpseudocode}
\usepackage{algorithmicx,algorithm}
\usepackage{booktabs}
\usepackage{multirow}
\usepackage{caption2}
\usepackage{cite}
\usepackage{float}
\usepackage{epstopdf}
\usepackage{subfigure}
\begin{document}
\title{{\LARGE \textbf{Bearing fault diagnosis based on domain adaptation using transferable features under different working conditions}}}
\author{Zhe Tong$^{1}$, Wei Li$^{1}$, Bo Zhang$^{2*}$, Meng Zhang$^{1}$\\
\\
$^{1}$School of Mechanical Engineering,\\
China University of Mining and Technology,\\
Xuzhou, 221116, P.R. China.\\
$^{2}$School of Computer Science and Technology,\\
China University of Mining and Technology,\\
Xuzhou, 221116, P.R. China. Email: zbcumt@163.com\\
}

\date{}
\maketitle
\doublespacing

\begin{abstract}
Bearing failure is the most common failure mode in rotating machinery and can result in large financial losses or even casualties. However, complex structures around bearing and actual variable working conditions can lead to large distribution difference of vibration signal between a training set and a test set, which causes the accuracy-dropping problem of fault diagnosis. Thus, how to improve efficiently the performance of bearing fault diagnosis under different working conditions is always a primary challenge. In this paper, a novel bearing fault diagnosis under different working conditions method is proposed based on domain adaptation using transferable features(DATF). The dataset of normal bearing and faulty bearings are obtained through the fast Fourier transformation(FFT) of raw vibration signals under different motor speeds and load conditions. Then we reduce marginal and conditional distributions simultaneously across domains based on maximum mean discrepancy(MMD) in feature space by refining pseudo test labels, which can be obtained by the Nearest-Neighbor(NN) classifier built on training data, and then a robust transferable feature representation for training and test domains is achieved after several iterations. With the help of the NN classifier trained on transferable features, bearing fault categories are identified accurately in final. Extensive experiment results show that the proposed method under different working conditions can identify the bearing faults accurately and outperforms obviously competitive approaches. \bigskip
\end{abstract}

\section{Introduction}
Bearings are the most critical components and widely used in rotating machinery, whose health conditions, for example, the fault degree in different places under different motor speeds and loads, may have huge effect on the performance, reliability and residual life of the equipment\cite{JACOBS2016144} or even can lead to heavy casualties\cite{LI2017356,873208,JALAN2009604}. Hence, it is important to diagnose bearings under different working conditions.\\
\indent Cracks or spalls on the surfaces of the roller, outer race or inner race are commonly failure modes in bearings\cite{QIAO201672}. Vibration signal is the most intuitive description for operating state of bearing. With the vibration signals under different conditions are collected by sensors\cite{LI201680}, many intelligent fault diagnosis methods have already achieved significant success in the field of fault diagnosis. In\cite{HUANG20111018}, a genetic algorithm-based SVM(GA-SVM) model was presented, and it had high accuracy and generalization ability by optimizing parameters of SVM. N. Saravanan et al \cite{SARAVANAN20104168} proposed fault diagnosis method based on DWT and ANN, and it has been proved such approach had the potential to diagnose various faults of the gear box. There are two key points for common intelligent fault diagnosis technologies, namely, feature extraction and classification. Raw vibration signal collected by sensors is abound in redundant information. Thus, it is important for fault diagnosis to achieve effective features\cite{s17020425}. Many signal processing approaches are applied to feature extraction from vibration signals. Such as, time-domain statistical analysis, frequency-domain analysis\cite{JOANNIN201775} and time-frequency domain analysis\cite{LI2017356}. Then reducing the dimensions is conducted for the sake of computational efficiency, such as principal component analysis(PCA)\cite{MISRA20021281}, locally linear embedding(LLE)\cite{Roweis2323} and linear discriminant analysis(LDA)\cite{6583974}. Finally, with the help of a suitable classifier, such as, nearest-neighbor (NN), support vector machine(SVM) or artificial neural networks(ANN), features acquired from above technological process are used for defect classification.\\
\indent To be true, most of intelligent fault diagnosis methods work well only under a general assumption: the training and test data are drawn from the same distribution. However, in operation of rotating machinery, because of complicated working conditions and complex sensor signals, the distribution of fault data is not consistent. Vibration signals sampled under different working conditions violate above assumption and show large distribution differences between domains\cite{s17020425,7961149}, which lead to drop dramatically of performance. More specifically, take the roller bearing fault diagnosis problem as an example, classifier was trained under a very concrete type of data sampled under a certain motor speed and load, however, the actual application in fault diagnosis is to recognize test data collected under another motor speed and load. Although the fault diameter and categories are not changed, the distribution differences between training data (training domain) and test data (test domain) changes with working condition varies. As a direct result, the classifier can achieve high accuracy on training domain while performing poorly on test domain\cite{7961149}. This is caused by distribution differences between two domains, since features extracted from one domain can not represent for another domain. Of course we can spend lots of time and efforts to recollect data to build a new classifier for effective fault diagnosis on test domain. However, we can not always to replace classifier by repetitively recollecting data. Worse, it is so expensive or even impossible to rebuild the fault diagnosis model from scratch using newly recollected training data for the actual task. Therefore, there is still plenty of room for improvement.\\
\indent In order to avoid such recalibration effort, we might want to refine a fault diagnosis model trained in one condition(training domain) for a new working condition(test domain), or to refine the model trained on one rolling bearing(training domain) for a new rolling bearing(test domain). This leads to the research of domain adaptation(DA)\cite{SHIMODAIRA2000227,7486184}. DA can be considered as particular setting of transfer learning\cite{5640675,5288526} which aims to leverage the knowledge learnt from a training domain to use in a different but related test domain by reducing distribution differences\cite{6550016,5288526}. Maximum mean discrepancy(MMD)\cite{7078994,7740016,Nigam:2000:TCL:347709.347724} in the field of DA can be applied to evaluate distribution divergences.\\
\indent In this paper, considering actual fault diagnosis application, we propose a novel bearing fault diagnosis under different working conditions based on domain adaptation using transferable features(DATF). Dataset of normal bearing and faulty bearings are achieved through the fast Fourier transformation(FFT) of raw vibration signals under different motor speeds and load conditions. Fault diagnosis model is built by using nearest-neighbor(NN) classifier in training domain, and then, we resort the pseudo outputs of NN classifier in test domain to refine this model by reducing distribution difference between domains constantly, so that transferable feature representation could be learnt from training and test domains . Finally, NN classifier is built with extracted transferable features and bearing faults are identified accurately.\\
\indent The rest of this paper is organized as follows. Section 2 sketches out previous works and preliminaries, including domain adaptation and maximum mean discrepancy. Section 3 introduces fault diagnosis using transferable features, including feature space generation and transferable feature extraction and diagnosis. Section 4 presents the experimental evaluations. The conclusion are given in Section 5.
\section{Previous works and preliminaries}
\subsection{Domain adaptation}
DA as one research of transfer learning aims at making full use of information coming from both training domain and test domain during the learning process to adapt automatically\cite{6550016,5288526,6126344}. Generally domain is considered as consisting of a feature space of inputs $\mathcal{X}$ and a probability distribution of inputs $P(X)$, where $X=\{x_1,\cdots,x_n\} \in \mathcal{X}$ is a series of learning samples. Note that distributions of two domains are diverse when source domain and target domain are different, that is $X_S \neq X_T$ and $P(X_S) \neq P(X_T)$\cite{7078994,Tahmoresnezhad2017}.\\
\indent In our work, the objective of domain adaptation is to extract transferable features between two domains for realizing successfully bearing fault diagnosis under different working conditions. We denote the labeled training domain $X_{tr}=\{(x_{tr_1},y_{tr_1}),...,(x_{tr_{n_1}},y_{tr_{n_1}})\}$, where $x_{tr_i} \in \mathcal{X}$ is the input and $y_{tr_i} \in \mathcal{Y}$ is the related class label. Similarly, let the unlabeled test domain be $X_{te}=\{(x_{te_1}),...,(x_{te_{n_2}})\}$, where the input $x_{te_i} \in \mathcal{X}$. In the aspect of distribution, let $P(X_{tr})$ and $Q(X_{te})$ be the marginal distributions of $X_{tr}=\{x_{tr_i}\}$ and $X_{te}=\{x_{te_i}\}$ from the training and test domains, respectively. Similarly let $P(Y_{tr} \vert X_{tr})$ and $Q(Y_{te} \vert X_{te})$ be the conditional distributions of $X_{tr}=\{x_{tr_i}\}$ and $X_{te}=\{x_{te_i}\}$ from the training domain and test domain, respectively\cite{7078994,DBLP:journals/corr/Csurka17,6751384}.\\
\indent In this literature, we focus on the following settings: 1)one training domain and one test domain share the same fault types and feature space. 2)domain adaptation in our work is unsupervised and training domain $X_{tr}$ are of labels while test domain $X_{te}$ are fully unlabeled. 3)the marginal distribution $P(X_{tr}) \neq Q(X_{te})$ and the conditional distribution $P(Y_{tr} \vert X_{tr}) \neq Q(Y_{te} \vert X_{te})$. Above settings are well suited to real-world variable working conditions fault diagnosis. Our task is predict the fault types of bearing accurately in the unlabeled test domain with entirely different distribution by using the model built in training domain.\\
\subsection{Maximum mean discrepancy}
Typical procedure of domain adaptation is to reduce marginal distribution difference across domains. In our work, domain adaptation is to reduce both marginal and conditional distribution difference simultaneously by explicitly minimizing the empirical distance measure, which is more suitable for the situation of bearing fault diagnosis under different working conditions. In order to void expensive distribution calculation caused by the parametric criteria, a nonparametric distance metric, known as MMD, is employed for domain adaptation in our work. Taking data from source domain $X_S$ and target domain $X_T$, the MMD calculates the empirical estimate of distances across domains in the $k$-dimensional embedding\cite{Tahmoresnezhad2017,7078994}:
\begin{equation}
  D_m(X_S,X_T)=||\frac{1}{n_s}\sum_{i=1}^{n_s}A^Tx_i-\frac{1}{n_t}\sum_{j=n_s+1}^{n_s+n_t}A^Tx_j||^2
\end{equation}
where $D_m$ is the distance of marginal distributions across domains, $A$ is the adaptation matrix, and $n_s$ and $n_t$ denote the number of source instances and target instances, respectively.\\
\section{Fault diagnosis using transferable features}
As mentioned in Section 1, huge distribution difference across training domain and test domain under different working conditions directly leads to poor performance of bearing fault diagnosis. In order to solve this problem, we need to learn the shift between two domains and extract more robust transferable features for two domains. In this section, we present our novel bearing fault diagnosis method under variable working conditions. The framework of our method is illustrated in Figure 1. As shown in Figure 1, fault diagnosis model built via labeled training data is iterated revision according to pseudo-label, and the final diagnostic results are obtained through above revised model. Details of each part are elaborated in the following subsections.
\begin{figure}[H]
  \centering
  \renewcommand{\captionlabeldelim}{\ }
  \centerline{\includegraphics[width=6in]{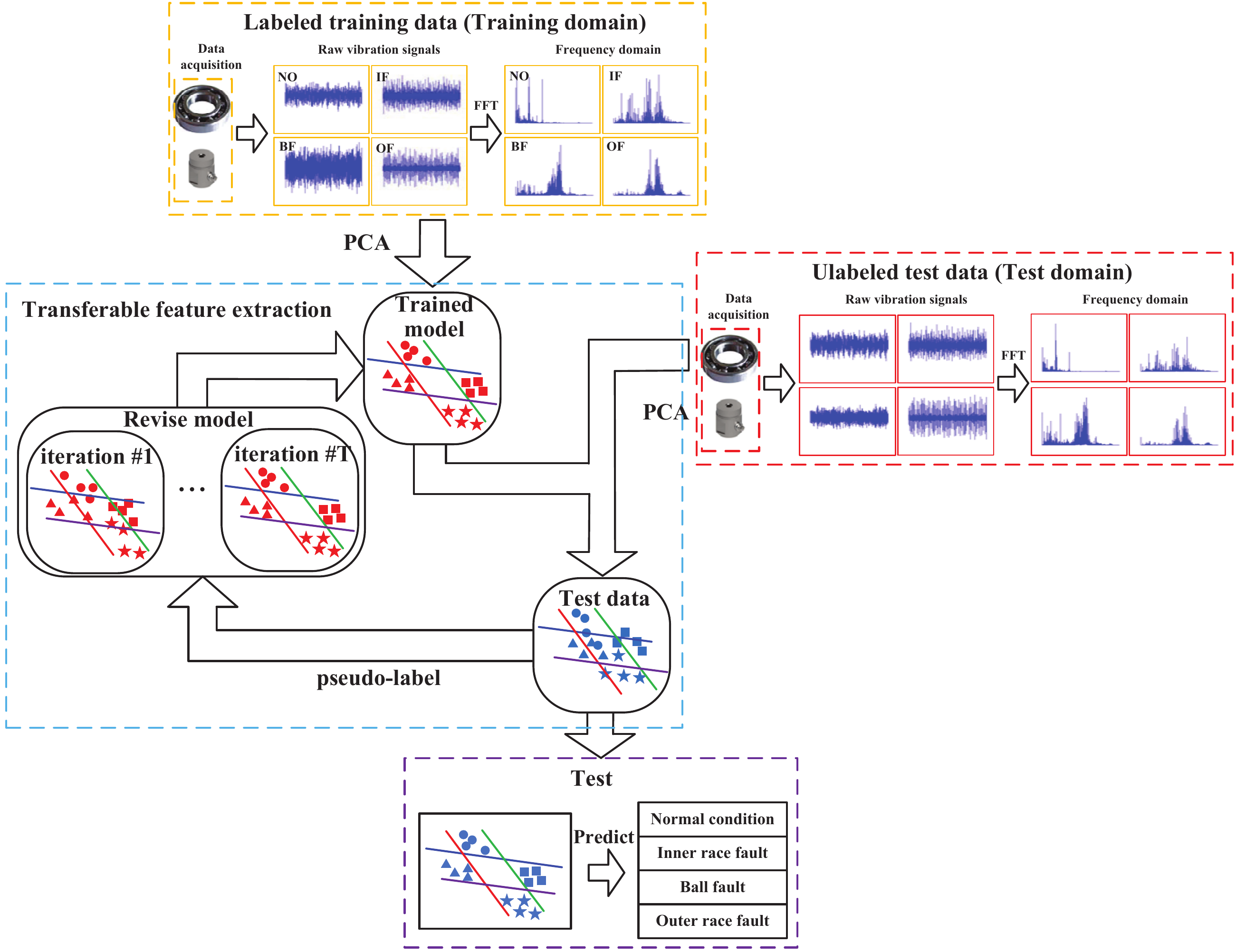}}
  \caption{The framework of DATF for variable working condition fault diagnosis}
\end{figure}
\subsection{Feature space generation}
Raw time series vibration signals are readily available and abound in bearing information. Owning to the rotating nature of raw vibration signals from a defective bearing, the periodic impulse would appear in obtained signals once a fault occurs. Thus, these fault impacts can be detected generally in frequency domain.\\
\indent In our work, we directly catch FFT amplitudes from the raw time series vibration signals as samples, where all samples have the same dimension, and these samples are generated under different motor speeds and load conditions, as described in figure 2.
\begin{figure}[H]
  \centering
  \renewcommand{\captionlabeldelim}{\ }
  \centerline{\includegraphics[width=3in]{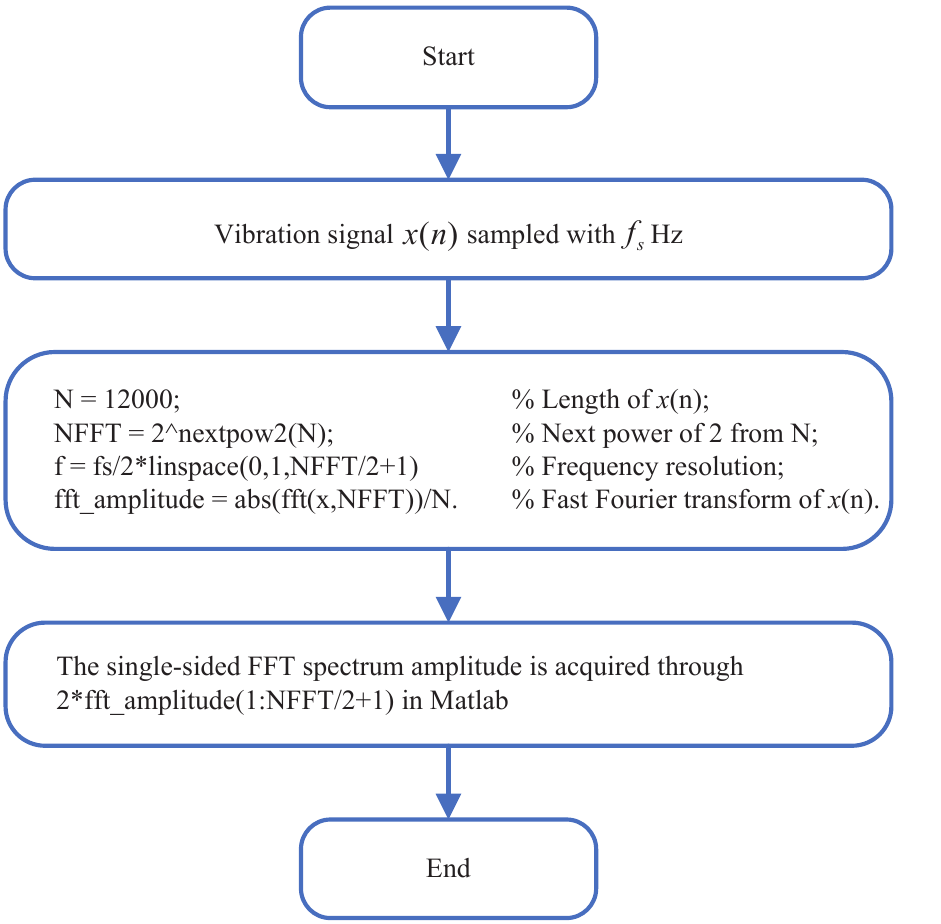}}
  \caption{Flowchart of FFT spectrum amplitudes creation in MATLAB}
\end{figure}
They are divided into two parts: labeled training data($D_{tr}$) and unlabeled test data($D_{te}$). Then we use principal component analysis(PCA) to generate feature space. The main steps of feature space generation are as follows:
\begin{itemize}
  \item {\bf Step 1}: Catch FFT amplitudes from raw time series vibration signals collected under different working conditions as samples $D_{data}$.
  \item {\bf Step 2}: Take one of the conditions with different fault types from $D_{data}$ as training samples $X_{tr} \in R^{n_{tr} \times d}$ with label $Y_{tr} \in R^{n_{tr} \times 1}$, and take another of the conditions with different fault types from $D_{data}$ as unlabeled test samples $X_{te} \in R^{n_{te} \times d}$.
  \item {\bf Step 3}: Denote $X_D = \{X_{tr},X_{te}\} \in R^{d \times (n_{tr} + n_{te})}$ and $H = I - \frac{1}{n_{tr} + n_{te}}ll^T$, where $I$ denotes the identity matrix and $l$ is considered as the ones vectors. Then, the $k$ dimensional representation is found by solving the following optimization problem $\underset{A^TA = I}{max} tr(A^TX_DHX_D^TA)$, and then, feature space is created by $V = A^TX_D$.
\end{itemize}
\subsection{Transferable feature extraction and diagnosis}
In order to reduce marginal distribution difference and extract robust feature for two domains, we resort MMD as the distance measures between $x_{tr}^i$ and $x_{te}^j$ to compare different distributions:
\begin{equation}
  ||\frac{1}{n_{tr}}\sum_{i=1}^{n_{tr}}A^Tx_i - \frac{1}{n_{te}}\sum_{j=n_{tr}+1}^{n_{tr} + n_{te}}A^Tx_j||^2 = tr(A^T X_D M_m X_D ^TA)
\end{equation}
where $M_m = \left [\begin{array}{cc} {(M_m)}_{tr,tr} & {(M_m)}_{tr,te} \\
{(M_m)}_{te,tr} & {(M_m)}_{te,te}
\end{array}\right]$ is the MMD matrix and is computed as follows\cite{6751384,Tahmoresnezhad2017}
\begin{equation}
M_m=\left\{
\begin{aligned}
&\frac{1}{n_{tr}n_{tr}}, \quad x_i, x_j \in X_{tr} \\
&\frac{1}{n_{te}n_{te}}, \quad x_i, x_j \in X_{te} \\
&\frac{-1}{n_{tr}n_{te}}, \quad otherwise
\end{aligned}
\right.
\end{equation}
The marginal distributions between training domain and test domain are brought closer under the new representation $V=A^TX_D$ by minimizing Eq.(2).\\
\indent In theory, training and test data under different working conditions collected from sensors should be of the same marginal and conditional distributions while the reality is very different. For improving the performance of bearing fault diagnosis under different work conditions, in our work, the differences of conditional distribution between domains are also reduced by mining the class-conditional distribution. Formally, the class-conditional distributions can be measured according to modified MMD.
\begin{equation}
  ||\frac{1}{n_{tr}}\sum_{i=1}^{n_{tr}}A^Tx_i - \frac{1}{n_{te}}\sum_{j=n_{tr}+1}^{n_{tr} + n_{te}}A^Tx_j||^2 = tr(A^T X_D M_c X_D^T A)
\end{equation}
Where $M_c = \left [\begin{array}{cc} {(M_c)}_{tr,tr} & {(M_c)}_{tr,te} \\
{(M_c)}_{te,tr} & {(M_c)}_{te,te}
\end{array}\right ]$ is MMD coefficient matrix that includes the class label $c$, and it can be calculated according to\cite{6751384,Tahmoresnezhad2017}
\begin{equation}
  M_c=\left\{
\begin{aligned}
&\frac{1}{n_{tr}^cn_{tr}^c}, \quad x_i, x_j \in X_{tr} \\
&\frac{1}{n_{te}^cn_{te}^c}, \quad x_i, x_j \in X_{te} \\
&\frac{-1}{n_{tr}n_{te}}, \quad \left\{
\begin{aligned}
&x_i \in X_{tr}^c, x_j \in X_{te}^c\\
&x_j \in X_{tr}^c, x_i \in X_{te}^c
\end{aligned}
\right.\\
&0, \quad otherwise
\end{aligned}
\right.
\end{equation}
The conditional distributions between training and test domains are brought closer under the new representation $V=A^TX_D$ by minimizing Eq.(4).\\
\indent In order to obtain effective and robust transferable feature representation and improve the quality of fault diagnosis, our work aims to reduce the impact of discrepancies from both the marginal and conditional distributions between training and test domains by resorting the pseudo labels of test data\cite{6751384} on diagnosis, and these pseudo labels can be obtained from a base classifier(NN classifier) built on the labeled training data to predict the fully unlabeled test data. Thus, the final optimization problem Eq.(6) in this paper is comprised from Eq.(2) and Eq.(4).
\begin{equation}
  \underset{A^TX_DHX_D^TA = I}{min} (1-\lambda)\sum_{c=0}^{C} tr(A^TX_DM_cX_D^TA) + \lambda ||A||_F^2
\end{equation}
where $||\cdot||_F$ is the Frobenius norm that guarantees the optimization problem to be well defined, and $\lambda$ is the regularization parameter\cite{Tahmoresnezhad2017} that trades off the impact of regularization term on the transformation matrix A. The goal is to find the latent feature space created by a transformation matrix $A \in R^{d \times k}$ where the discrepancies of both the marginal and conditional distributions between domains are significantly reduced. The Lagrange function for Eq.(7) is constructed, where $\Lambda = diag(\Lambda_1,\cdots,\Lambda_k) \in R^{k \times k}$ is the Lagrange multiplier.
    \begin{equation}
    L = (1-\lambda)tr(A^T(X_D\sum_{c=0}^{C}M_cX_D^T)A) + \lambda tr(A^TA) + tr((I - A^TX_DHX_D^TA)\Lambda)
    \end{equation}
According to $\frac{dL}{dA}=0$, the optimal solution of Eq.(9) can be acquired through the generalized eigen decomposition.
    \begin{equation}
    ((1-\lambda)X_D\sum_{c=0}^{C}M_cX_D^T + \lambda I)A = X_DHX_D^TA\Lambda
    \end{equation}
Finally, the adaptation matrix A is obtained from solving Eq.(8) for $k$ smallest eigenvectors. The procedure of fault diagnosis using DAFT can be depicted as follows in details:
\begin{itemize}
  \item {\bf Step 1}: For given training data $X_{tr} \in R^{n_{tr} \times d}$ with label $Y_{tr} \in R^{n_{tr} \times 1}$ and unlabeled test data $X_{te} \in R^{n_{te} \times d}$ in the feature space.
  \item {\bf Step 2}: Construct MMD matrix $M_m$ by Eq.(2). Adaptation matrix $A$ generated by the $k$ smallest eigenvectors can be acquired by solving Eq.(8) through Lagrange multiplier. Then the robust representation for two domains is obtained $V = A^TX_D$.
  \item {\bf Step 3}: Train the NN classifier on projected training data$\{A^TX_{tr}, Y_{tr}\}$, and then obtain pseudo test data labels $Y_{te}$ that denote the conditional probability $Q(Y_{te} \vert X_{te})$ by using the trained NN classifier.
  \item {\bf Step 4}: Update MMD matrix $\{M_c\}_{c=1}^C$ by Eq.(5) according to $P(Y_{tr} \vert X_{tr}) = Q(Y_{te} \vert X_{te})$, and then obtain the updated adaptation matrix $A$ by solving Eq.(8) through Lagrange multiplier. The updated robust representation for two domains is obtained $V = A^TX_D$, and then jump to Step 3 until the end of the iteration.
  \item {\bf Step 5}: Finally the test data labels $Y_{te}$ are predicted accurately by the adaptive NN classifier.
\end{itemize}
\section{Experimental evaluations}
In order to demonstrate the effectiveness of the proposed fault diagnosis method, the vast bearing vibration signals collected from a bearing test rig are used. Dataset is acquired from the bearing data centre of Case Western Reserve University(CWRU)\cite{Case-Western-Reserve-University}. DATF is compared with the baseline approaches and several successful methods.\\
\indent a. Baseline: NN classifier with no projection and no adaptation is created. That is, original input is directly used for diagnosis.\\
\indent b. NN NA: NN classifier with no adaptation is created. Specifically, we use a new representation extracted from original input by PCA without domain adaptation.\\
\indent c. NN SA: NN classifier with projection and domain adaptation  using subspace alignment that only reduces the marginal distribution\cite{6751479}.\\
\indent a is a baseline method without projection and domain adaptation techniques, which is widely used in the field of fault diagnosis. b is a classical method without domain adaptation, which has achieved success in many fault diagnosis applications. c is one of the novel and efficient approach in domain adaptation.
\subsection{Experimental setup and dataset preparation}
The test-bed illustrated in figure 3 consists of a driving motor, a 2 hp motor for loading, a torque sensor/encoder, a power meter, accelerometers and electronic control unit\cite{ALBUGHARBEE2016246,Case-Western-Reserve-University}. The test bearings locate in the motor shaft. Subjected to electrosparking, inner-race faults (IF), outer-race faults (OF) and ball fault (BF) of different sizes (0.007in, 0.014in, and 0.021in) are introduced into the drive-end bearing of motor\cite{0957-0233-27-3-035005}. The vibration signals are sampled with the help of accelerometers installed to the rack with magnetic bases.\\
\begin{figure}[H]
  \centering
  \renewcommand{\captionlabeldelim}{\ }
  \centerline{\includegraphics[width=3in]{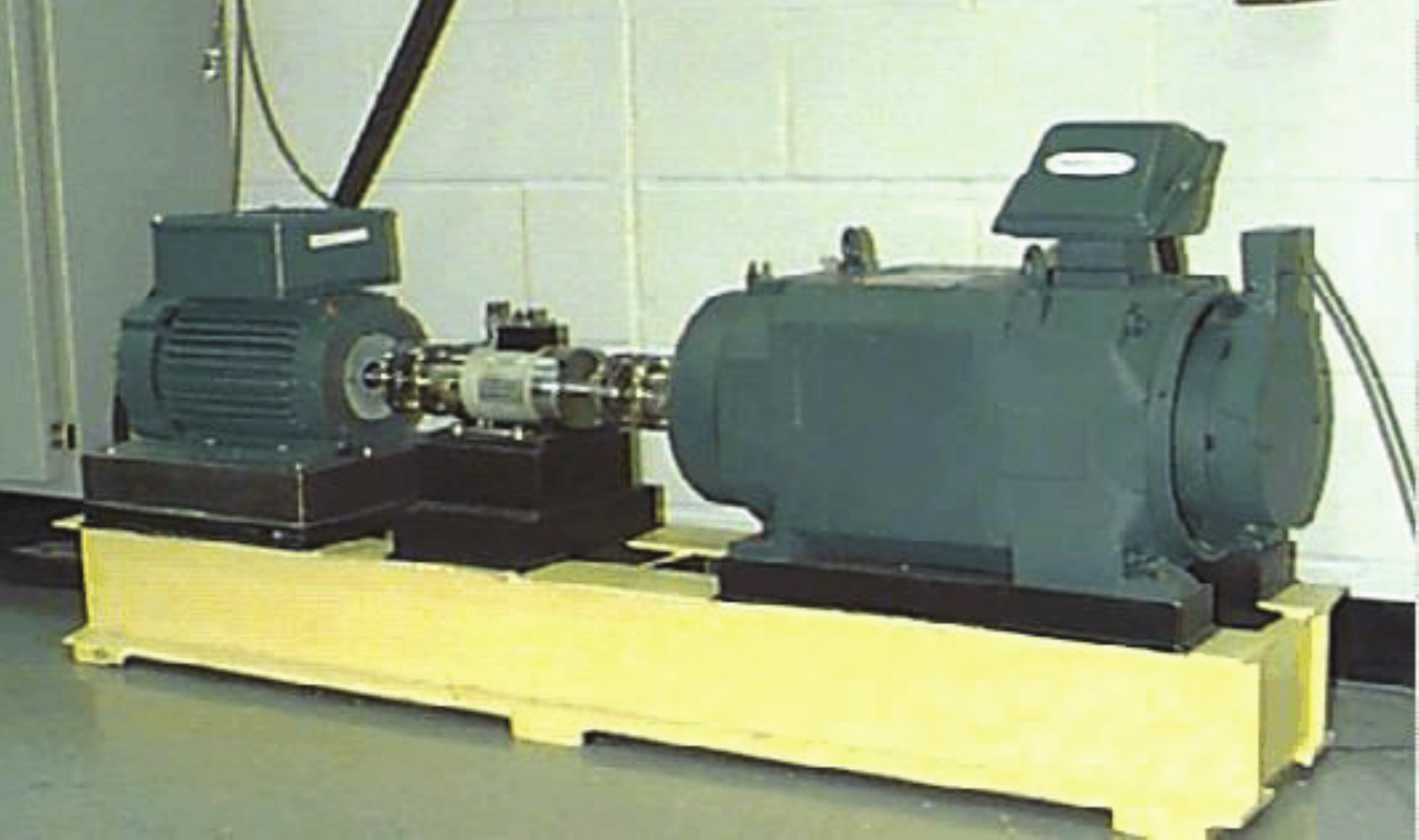}}
  \caption{Bearing test rig of Case Western Reserve University Data Center}
\end{figure}
\indent The working condition of the rotating machinery is usually complex in real-world. For purpose of simulating the actual application and making the experimental results more persuasive, in our experiment, dataset, collected from Drive End Bearing Fault Data and sampled at a frequency of 12kHz, is obtained from different working conditions. Dataset includes three kinds of fault degrees(0.007in, 0.014in and 0.021in). Each fault degree contains four fault types of bearings: NO, IF, OF and BF. Each fault type of vibration data is collected from four kinds of working conditions, i.e., L0 = 0 hp/1797 rpm, L1 = 1 hp/1772 rpm, L2 = 2 hp/1750 rpm and L3 = 3 hp/1730 rpm. Each sample contains 2049 Fourier coefficients transformed from the raw vibration signals using FFT. Each domain on dataset contains four fault types and each fault type contains 200 samples. Under our experimental setup, it is impossible to find the optimal $k$ and $\lambda$ via cross validation, since labeled training data and unlabeled test data are sampled from different working conditions. Thus, empirically searching the parameter space is used to find the optimal parameter settings, and details are described in Section 4. Finally, $\lambda = 0.1$ and $k = 100$ are used in our work.\\
\indent In order to verify the benefits of DATF, contrast methods of a-c are also carried out simultaneously. The scenario settings of all experiments are trained on labeled training data under one single load(training domain) to diagnose the unlabeled test data under another load(test domain). In all, 48 different transferring tests are conducted and the description of experimental setup in detail is shown in Table 1.\\
\begin{table}[htbp]
\begin{center}
\renewcommand{\captionlabeldelim}{\ }
  \caption{Description of the experimental setup}
  \begin{tabular}{ccccc}
    \toprule
    \multicolumn{1}{c}{Task} & \multicolumn{4}{c}{Diagnose unlabeled test samples in test domain}\\
    \cline{2-5}
    $\sharp$ of & Labeled training & Unlabeled test & Fault & Fault\\
     test & (training domain) & (test domain) & type & size \\
    \midrule
    1 & L0,L1,L2,L3 & L0 & NO,IF, & 0.007in \\
      &       &                         & BF,OF  &          \\
    2 & L0,L1,L2,L3 & L1 & NO,IF, & 0.007in \\
      &       &                         & BF,OF  &          \\
    3 & L0,L1,L2,L3 & L2 & NO,IF, & 0.007in \\
      &       &                         & BF,OF  &          \\
    4 & L0,L1,L2,L3 & L3 & NO,IF, & 0.007in \\
      &       &                         & BF,OF  &          \\
    5 & L0,L1,L2,L3 & L0 & NO,IF, & 0.014in \\
      &       &                         & BF,OF  &          \\
    6 & L0,L1,L2,L3 & L1 & NO,IF, & 0.014in \\
      &       &                         & BF,OF  &          \\
    7 & L0,L1,L2,L3 & L2 & NO,IF, & 0.014in \\
      &       &                         & BF,OF  &          \\
    8 & L0,L1,L2,L3 & L3 & NO,IF, & 0.014in \\
      &       &                         & BF,OF  &          \\
    9 & L0,L1,L2,L3 & L0 & NO,IF, & 0.021in \\
      &       &                         & BF,OF  &          \\
    10 & L0,L1,L2,L3 & L1 & NO,IF, & 0.021in \\
       &       &                         & BF,OF  &          \\
    11 & L0,L1,L2,L3 & L2 & NO,IF, & 0.021in \\
       &       &                         & BF,OF  &          \\
    12 & L0,L1,L2,L3 & L3 & NO,IF, & 0.021in \\
       &       &                         & BF,OF  &          \\
    \bottomrule
  \end{tabular}
  \end{center}
\end{table}

\subsection{Diagnosis results of the proposed method}
The diagnositic results for fault size being 0.007in, 0.014in and 0.021in are shown in figure 4, figure 5 and figure 6. The average classification accuracies of four methods are described in figure 7.
\begin{figure}[H]
  \centering
  \renewcommand{\captionlabeldelim}{\ }
  \centerline{\includegraphics[width=5in]{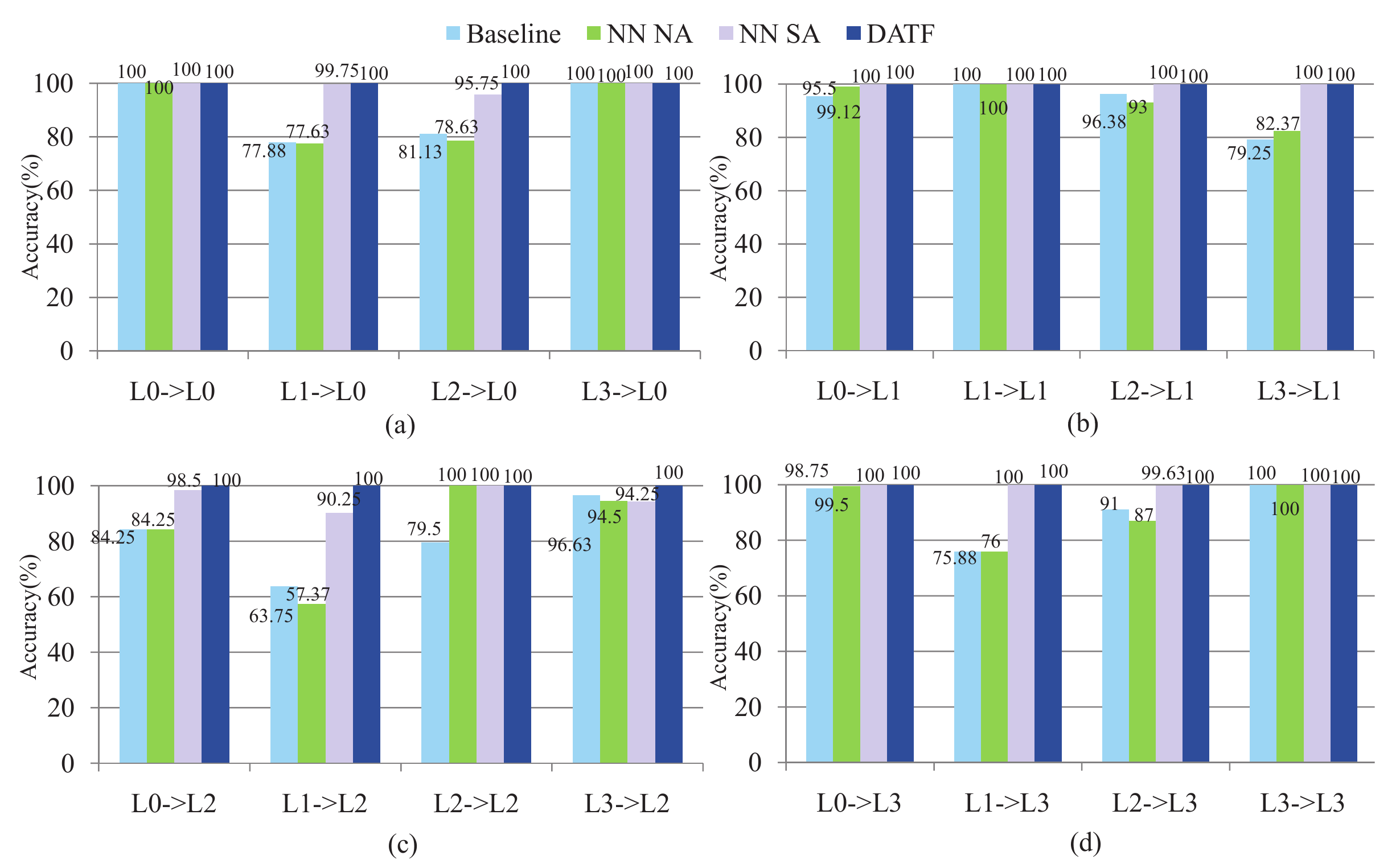}}
  \caption{The results with fault size being 0.007in}
\end{figure}

\begin{figure}[H]
  \centering
  \renewcommand{\captionlabeldelim}{\ }
  \centerline{\includegraphics[width=5in]{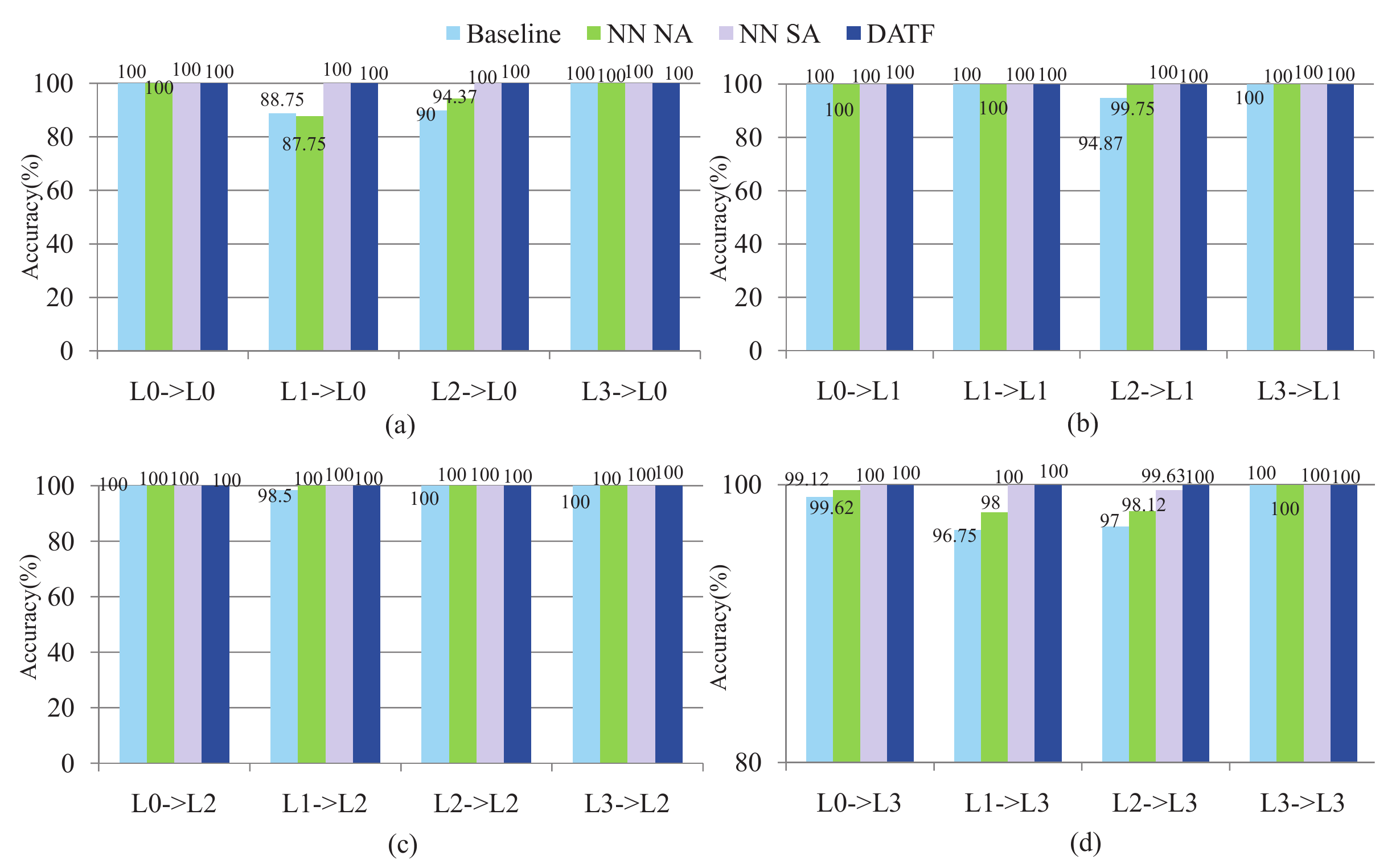}}
  \caption{The results with fault size being 0.014in}
\end{figure}

\begin{figure}[H]
  \centering
  \renewcommand{\captionlabeldelim}{\ }
  \centerline{\includegraphics[width=5in]{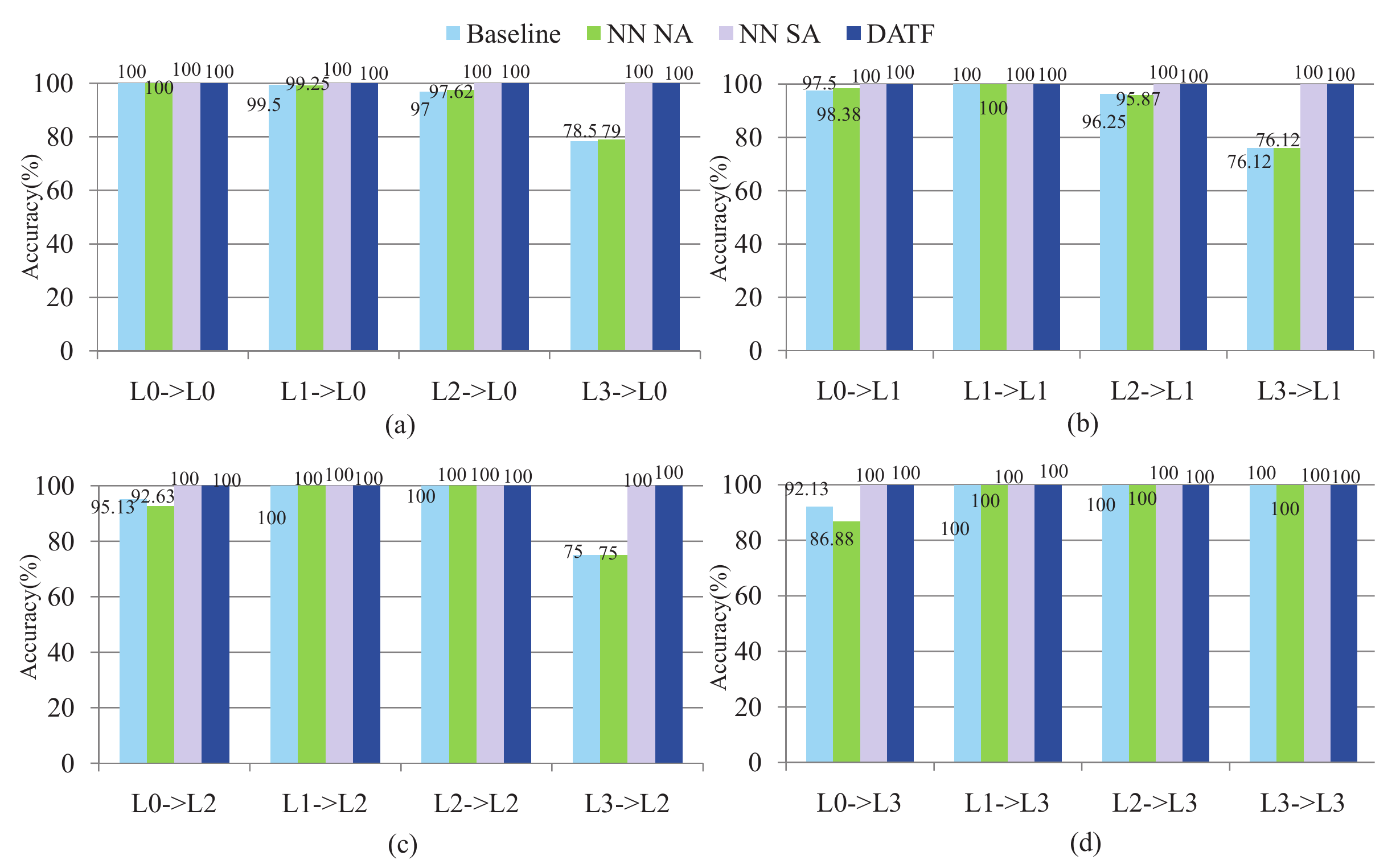}}
  \caption{The results with fault size being 0.021in}
\end{figure}

\begin{figure}[H]
  \centering
  \renewcommand{\captionlabeldelim}{\ }
  \centerline{\includegraphics[width=5in]{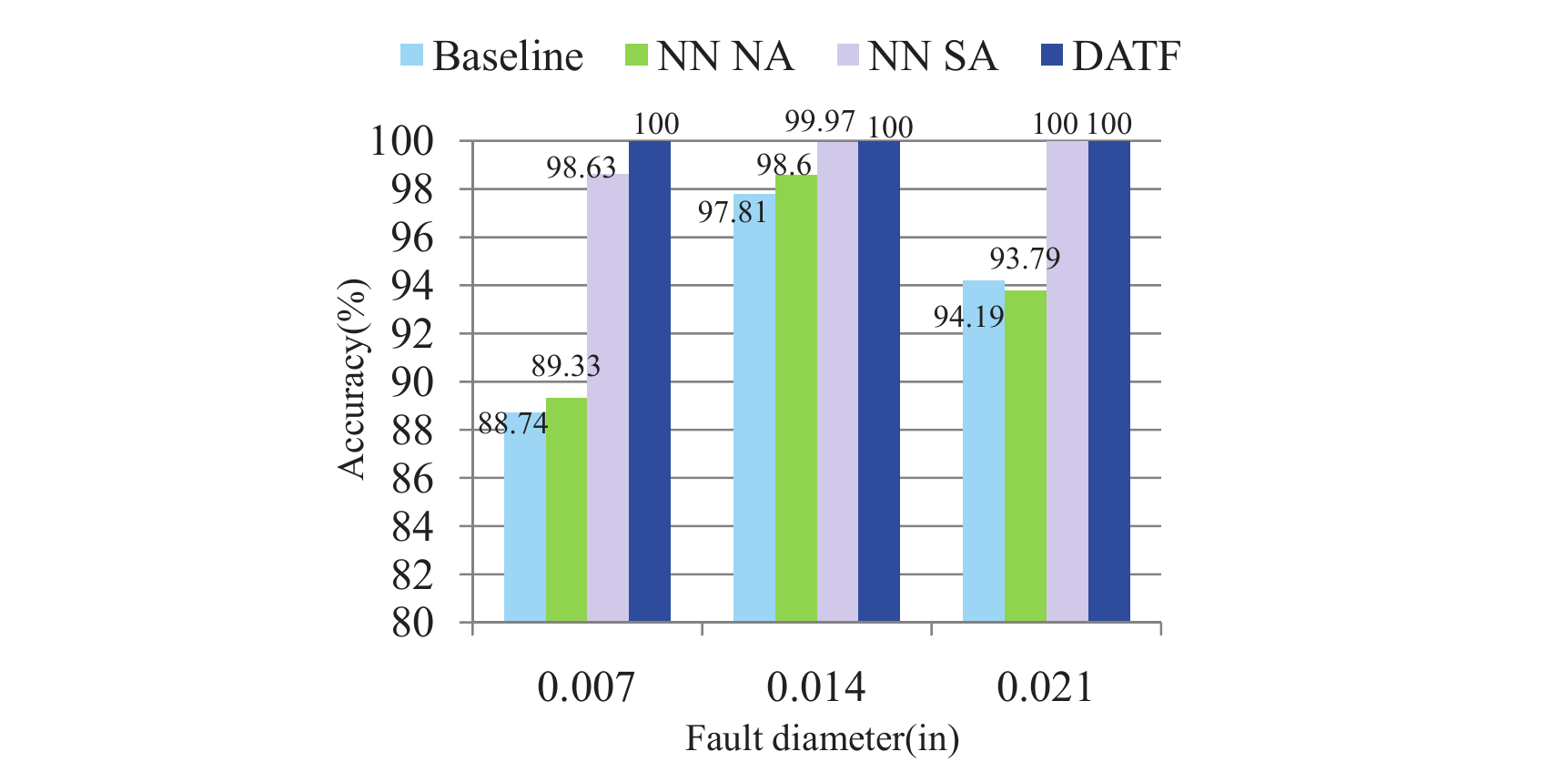}}
  \caption{The average classification accuracies}
\end{figure}

Each figure is composed of four subfigures and test domains in every figure are ordered clockwise from the top left: L0, L1, L2 and L3. The left of the symbol "$->$" in every subfigures represents the training domain and the right represents the test domain. For each set of bars in figure 4, 5 and 6, the performances indicate transferring from training domain to test domain, which simulates fault diagnosis under different working conditions. The load and speed between different domains have large discrepancies. For example, in figure 4(a), the test domain is L0(the motor load is 0hp and speed is 1797rpm), the training domain are L1(the motor load is 1hp and speed is 1772rpm), L2(the motor load is 2hp and speed is 1750rpm) and L3(the motor load is 3hp and speed is 1730rpm).\\
\indent From the performances of bearing fault diagnosis in figure 4, 5 and 6, the highest accuracy rates can always be achieved when the training set of one domain is the same with the testing set of one domain and this phenomenon is reasonable theoretically. We can obviously find that performances of the baseline method and NN NA are all very poor. For example, in figure 6(a), (b), (c), the accuracies are only about $75\%$ when we transfer L3 to L0, L1 and L2 respectively. Especially in figure 4, a lot of accuracies of baseline method and NN NA can not reach $70\%$ when we transfer L1 to L2. These results illustrate traditional methods without domain adaptation can not be applied to fault diagnosis in variable working conditions. The performances of NN SA are better than the first two types methods. In figure 5 and 6, the accuracies of NN NA for variable working condition bearing fault diagnosis are very high. However, in figure 4(c), the performance that transferring between L1 and L2 is only about $90\%$ and the accuracy is about $94\%$ when we transfer L3 to L2. Similar phenomena also appear in figure 4(a). These results mentioned above indicate that NN NA also can not be applied to complex and variable working condition bearing fault diagnosis. What is exciting that the proposed method is evidently superior to the other three compared methods in all cases, whatever the training domain and test domain are. Note that the accuracies of DATF all can achieve $100\%$ in figure 4, 5 and 6. Even in figure 4(a), DATF can still achieve a favorable accuracy($100\%$) while baseline method and NN NA just reach about $60\%$ and NN SA only achieve $90\%$ when transferring from L1 to L2. Compared to the other three methods, the average classification accuracy($100\%$) of DATF has been markedly improved. These results are all obtained from the benchmark datasets of fault diagnosis research under a relatively fair experiment condition. Through above result analysis, we can conclude that the proposed method is very potential for solving bearing fault diagnosis problems under different working conditions.\\
\indent To further illustrate the influence of extracted transferable features on the results, receiver operating characteristics (ROC) is applied to evaluation\cite{LEE1997637}. An ROC curve is generated by plotting the false positive rate and true positive rate as the threshold level is varied. In this paper, ROC curves are obtained from different models based on NN classifier, which are built on different extracted features, and we only report ROC results on transferring test that transfers L1 to L2 with fault size being 0.007in in figure 8, while similar trends on all other tests. Before the iteration begins in figure 8(a), performances of the model built on extracted features are unsatisfactory. After iteration 1 time in figure 8(b), performances of the model built on extracted transferable features are improved dramatically, and what is exciting is that performances based on extracted transferable features achieve the perfect detection results ultimately.
\begin{figure}[H]
  \centering
  \renewcommand{\captionlabeldelim}{\ }
  \centerline{\includegraphics[width=7in]{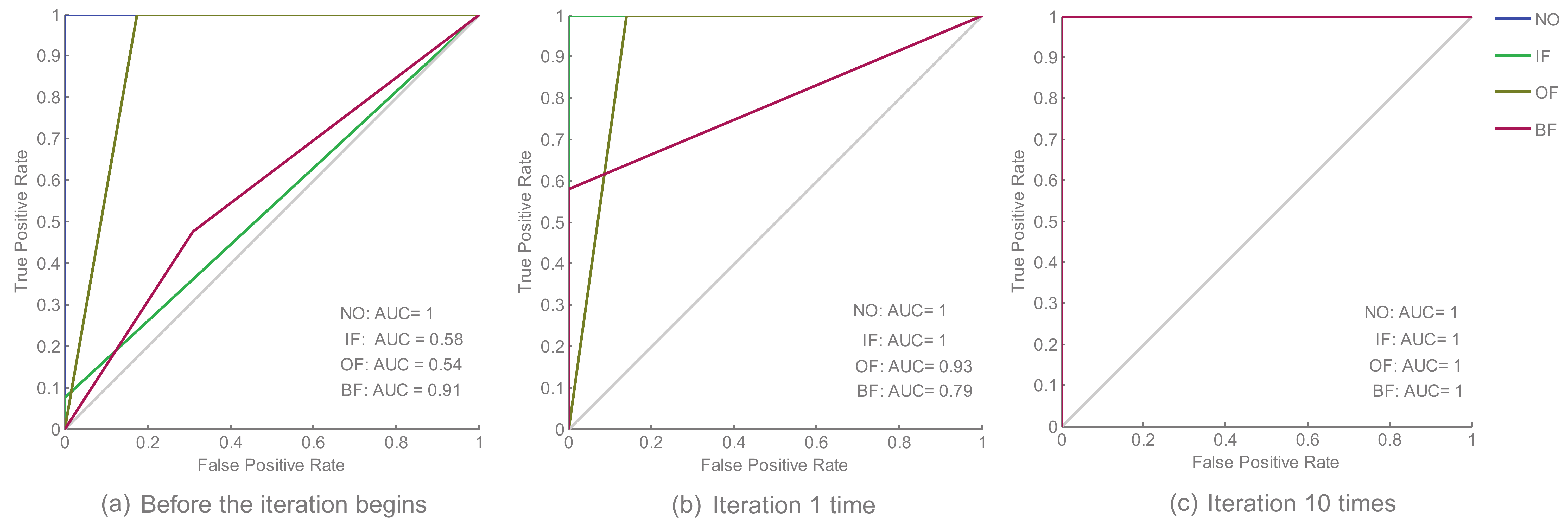}}
  \caption{ROC curves of faults detection based on DATF}
\end{figure}

\subsection{Parameter sensitivity}
In this section, we investigate the influence of the parameter $\lambda$, which represents regularization parameter and feature dimensionality respectively during transferable feature extraction. Theoretically, larger values of $\lambda$ can make shrinkage regularization more important in our work. When $\lambda \rightarrow 0$ and $\lambda \rightarrow 1$, the optimization problem is ill-defined. Different $\lambda$ has different effects on classification accuracy. Figure 9 reports the results. From the figure 9, it is obvious that different $\lambda$ have a great influence on diagnostic results with fault size being 0.007in and performances with fault size being 0.021in and it has little overall effect on results with fault size being 0.014in. What is noticeable is that results are little affected by parameter $\lambda$ when the training domain and test domain are the same, and $\lambda \in $ [0.05,0.5] can be optimal parameter values, which can indicate the proposed method can achieve stable and excellent performance under a wide range of parameter values.
\begin{figure}[H]
  \centering
  \renewcommand{\captionlabeldelim}{\ }
  \centerline{\includegraphics[width=7in]{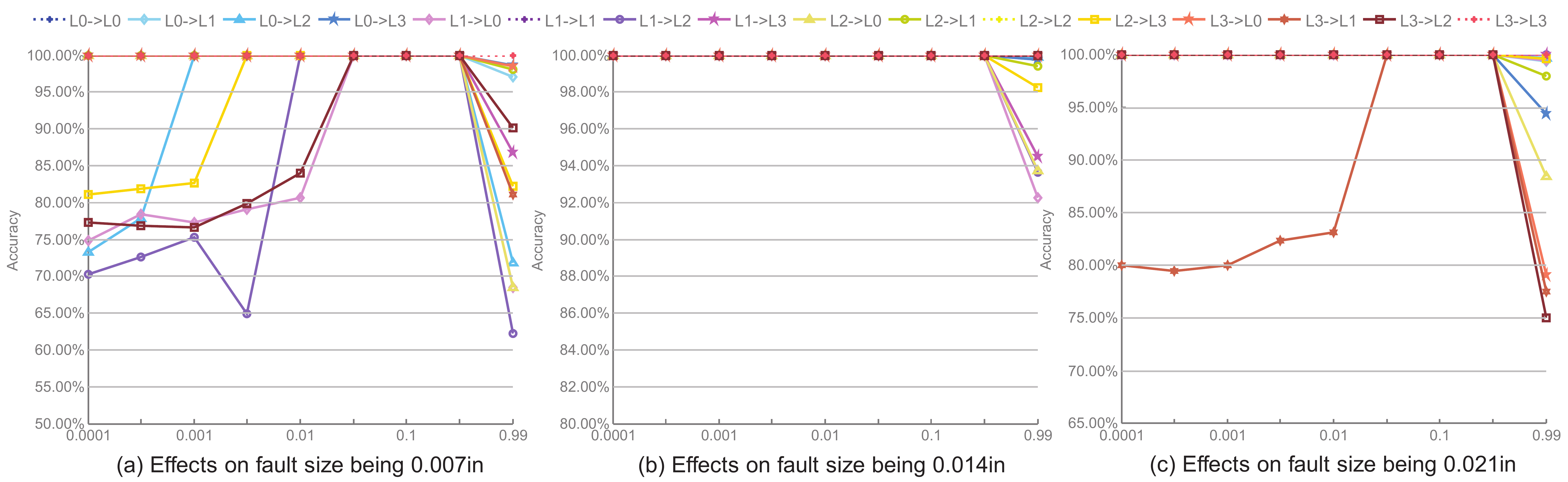}}
  \caption{Accuracy (\%) on different $\lambda$}
\end{figure}

\subsection{Domain discrepancy effect of empirical analysis}
In many actual fault diagnosis and classification scenarios, the distribution of training data domain is different from the testing data domain, which leads to fault diagnostic accuracy-dropping. In fact, the data distribution differences between domains(training data domain and test data domain) reflect the differences of the data structures that contain plenty of fault messages. It is a key point for fault diagnosis to extract fault features from data structures. In order to profound understand the effect of distribution differences between two domains and explain why the proposed method works, we resort the t-SNE technique\cite{LVDMaaten} to visualize high dimensional representation of mentioned methods in our experiment in a two-dimensional map.\\
\indent In all above mentioned cases, taking the transferring test that transfers L1 to L2 with fault size being 0.007in as an example in figure 10.
\begin{figure}[H]
  \centering
  \renewcommand{\captionlabeldelim}{\ }
  \centerline{\includegraphics[width=7in]{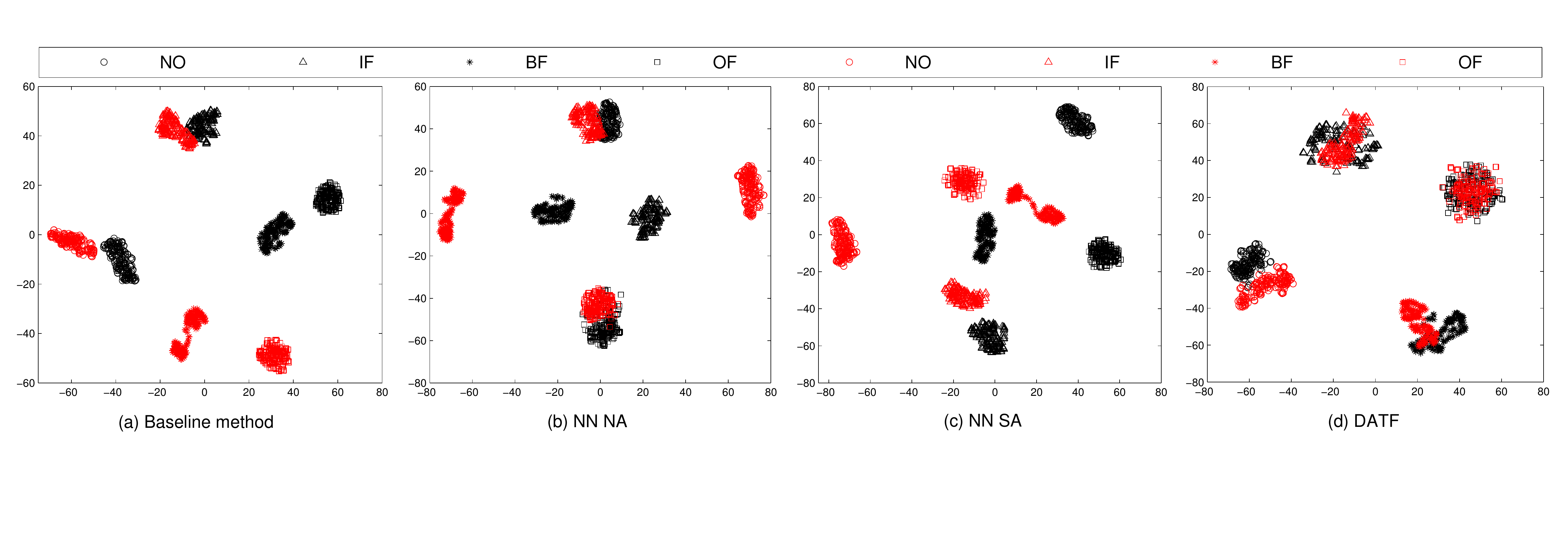}}
  \caption{Feature visualization via t-SNE\cite{LVDMaaten} over a fault diagnosis task from training domain L1(blue) to test domain L2(red) under different working conditions}
\end{figure}
From figure 10, it is clear that the distribution discrepancies of transferable features extracted via DATF between training domain and test domain are much smaller than the compared methods', and transferable features are much more divisible than others'. These results verify that DATF can figure out a robust feature representation for training domain and test domain, and test samples can be discriminated significantly with NN classifier built in training domain by using extracted transferable features.
\subsection{Discussion}
The proposed method provides a way of domain adaptation to extract robust fault features and classify fault types under different working conditions. Several remarks still need to be described.\\
(1) This work presents a new point of view that uses domain adaptation to realize bearing fault diagnosis under different working conditions. Li\cite{0957-0233-27-3-035005} utilized spectrum images as features to conduct bearing fault diagnosis, which applied two-dimensional principal component analysis (2DPCA) into the dimension reduction of the spectrum images of vibration signals and feature extraction, and most accuracies were very high. Unfortunately, there are still several instances having lower accuracies. To solve this problem, we apply the domain adaptation into this field and transferable features for training domain and test domain are extracted to classify fault types. Finally the accuracies all can reach $100\%$. In this paper, our work considers more bearing conditions(fault size being 0.007in). Compared with the method\cite{0957-0233-27-3-035005} in this situation, advantages of our method are highlighted.\\
(2) The vast results indicate that the proposed method is suitable for effectively classifying mechanical health conditions under different working conditions. In \cite{s17020425}, Deep Convolutional Neural Networks with Wide First-layer Kernel (WDCNN) and AdaBN are applied to diagnose three datasets which contain 10 kinds of health conditions (BF IF OF with fault size being 0.007 in, 0.014 in and 0.021 in) under three load conditions (Load1, Load2, Load3), respectively, which is similar to L1, L2 and L3 in this paper. The average accuracy of this method in \cite{s17020425} is $95.9\%$, whereas average accuracy of DATF is $100\%$. The main reason is that transferable features extracted based on domain adaptation take full advantage of structure information of training domain and test domain, and the distributions of transferable features extracted from training domain and testing domain are very close after our methods as shown in figure 10.\\
(3) It is noted that our method is unsupervised and focuses on fault transfer diagnosis based on the same fault diameter under different working conditions.
In \cite{7961149}, a method based on Neural Network by using transferring parameters is proposed and success for diagnosing two datasets including 6 kinds of health conditions which sampled from different fault diameters (BF IF OF with fault size being 0.007 in and 0.021 in) with the same motor load and speed (L0), and it focuses on fault diagnosis between two kinds of fault diameters under the same working conditions. In addition, unlike our method, it should be noted that a small amount of labeled data in test domain are needed when training modified neural networks, while our method does not need labeled test data during the training.

\section{Conclusion}
This paper presents a new way for solving bearing fault diagnosis under different working conditions. Although baseline approaches and several successful methods are all capable of detecting the bearing defects, distributional difference of datasets sampled from different working conditions has a huge impact on these methods, and their shallow representations are insensitive to distinguish different patterns under different working conditions. To tackle this problem, DATF extracts transferable feature representation for training and test domain by reducing the discrepancy between domains and strengthen the recognizable information in raw vibration signal. To evaluate the proposed DATF method, bearing fault diagnosis experiments were carried out. Extensive experiment results show DATF is capable of improving the performance of bearing fault diagnosis under different working conditions, comparing with the peer methods.

\begin{spacing}{1.5}
\textbf{Acknowledgements}
\end{spacing}

This research is supported by National Key R\&D Program of China (2016YFC0802900), National Natural Science Foundation of China (No. 51475455), the Natural Science Foundation of Jiangsu Province (No. BK20160276).\\

\bibliography{Finalref}

\end{document}